\documentclass{article}

\usepackage{microtype}
\usepackage{graphicx}
\usepackage{subfigure}
\usepackage{booktabs} 
\usepackage{multirow}
\usepackage{threeparttable}

\usepackage{hyperref}

\usepackage[accepted]{icml2025}

\usepackage{amsmath}
\usepackage{amssymb}
\usepackage{mathtools}
\usepackage{amsthm}

\usepackage[capitalize,noabbrev]{cleveref}

\theoremstyle{plain}

\theoremstyle{definition}

\theoremstyle{remark}

\usepackage[textsize=tiny]{todonotes}

\icmltitlerunning{Layer Separation: Adjustable Joint Space Width Images Synthesis in Conventional Radiography}

\begin{document}

\twocolumn[
\icmltitle{Layer Separation: Adjustable Joint Space Width Images Synthesis in Conventional Radiography}



\icmlsetsymbol{equal}{*}


\begin{icmlauthorlist}
\icmlauthor{Haolin Wang}{a}
\icmlauthor{Yafei Ou}{b}
\icmlauthor{Prasoon Ambalathankandy}{c}
\icmlauthor{Gen Ota}{d}
\icmlauthor{Pengyu Dai}{b}
\icmlauthor{Masayuki Ikebe}{d}
\icmlauthor{Kenji Suzuki}{b}
\icmlauthor{Tamotsu Kamishima}{e}
\end{icmlauthorlist}

\icmlaffiliation{a}{Graduate School of Health Sciences, Hokkaido University, Sapporo, Japan}
\icmlaffiliation{b}{Institute of Integrated Research, Institute of Science Tokyo, Yokohama, Japan}
\icmlaffiliation{c}{Processor Research Team, RIKEN Center for Computational Science, Kobe, Japan}
\icmlaffiliation{d}{Research Center For Integrated Quantum Electronics, Hokkaido University, Sapporo, Japan}
\icmlaffiliation{e}{Faculty of Health Sciences, Hokkaido University, Sapporo, Japan}

\icmlcorrespondingauthor{Yafei Ou}{ou.y.ac@m.titech.ac.jp}

\icmlkeywords{Machine Learning, ICML}

\vskip 0.3in
]



\printAffiliationsAndNotice{}  

\begin{abstract}
Rheumatoid arthritis (RA) is a chronic autoimmune disease characterized by joint inflammation and progressive structural damage. Joint space width (JSW) is a critical indicator in conventional radiography for evaluating disease progression, which has become a prominent research topic in computer-aided diagnostic (CAD) systems.
However, deep learning-based radiological CAD systems for JSW analysis face significant challenges in data quality, including data imbalance, limited variety, and annotation difficulties. 
This work introduced a challenging image synthesis scenario and proposed Layer Separation Networks (LSN) to accurately separate the soft tissue layer, the upper bone layer, and the lower bone layer in conventional radiographs of finger joints. 
Using these layers, the adjustable JSW images can be synthesized to address data quality challenges and achieve ground truth (GT) generation.
Experimental results demonstrated that LSN-based synthetic images closely resemble real radiographs, and significantly enhanced the performance in downstream tasks. 
The code and dataset will be available.
\end{abstract}

\section{Introduction}

\begin{figure}[!t]
    \centering
    \includegraphics[width=\linewidth]{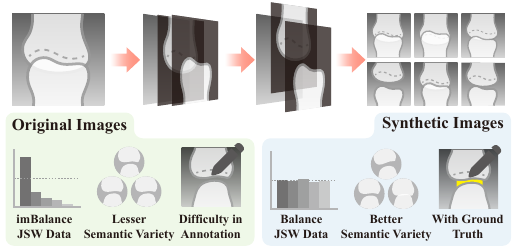}
    \caption{The adjustable JSW synthetic images are generated by producing layer images, following random shifting of the bone layers, and reconstruction with soft tissue layer. \textbf{Original Images}: imbalanced distribution of JSW, limited semantic variety, and difficulty in manual annotation. \textbf{Synthetic Data}: balanced distribution, enhanced semantic variety, and generative ground truth (GT) annotations.}
    \label{fig:background}
\end{figure}

Rheumatoid arthritis (RA) is a chronic autoimmune inflammatory disease characterized by joint swelling and tenderness, resulting in progressive joint destruction combined with severe disability. 
In the diagnosis and management of RA, radiographic analysis plays a crucial role, and changes in joint space width (JSW) are recognized as a key indicator for assessing and monitoring disease progression ~\cite{aletaha2018diagnosis, platten2017fully}. However, conventional radiographic analysis relies heavily on the expertise and judgment of radiologists, which is limited by subjectivity, leading to low accuracy and sensitivity. Therefore, the development of computer-aided diagnostic (CAD) methods is considered urgent and significant ~\cite{kingsmore2021introduction, stoel2024deep}. Deep learning-based CAD methods in joint space narrowing (JSN) progression quantification \cite{ou2022sub,wang2023deep}, JSW quantification \cite{langs2008automatic}, and Sharp/van der Heijde (SvdH) scoring \cite{hirano2019development}, are critically dependent on annotated data from experienced radiologists.
Nevertheless, publicly available datasets with comprehensive annotations are scarce, while private datasets referenced in existing studies are typically small (often limited to 100 - 200 conventional radiographic images)~\cite{stoel2024deep, ahalya2022automated}.
Additionally, the existing datasets are further exacerbated by insufficient variety, inconsistent imaging quality, and significant imbalances in JSW distribution (early-stage RA samples are much larger than late-stage samples), as shown in Figure \ref{fig:background}.
Meanwhile, the limitations of conventional radiography and the complexity of the joint structures pose significant challenges for accurate joint annotation (joint classification and mask labeling).
These limitations in annotated data critically hinder the performance of deep learning models and reduce the applicability of advanced models that require large datasets.

To address these data challenges, synthetic data has recently emerged as a promising approach in medical imaging~\cite{koetzier2024generating}. By generating image datasets that encompass diverse pathological features, varying levels of severity, and different regions, synthetic data enriches the data information available for model training. Synthetic data effectively tackles challenges related to limited patient populations, inconsistent data quality, and imbalanced distributions of disease stages. In addition, synthetic data mitigates biases introduced during data collection~\cite{paproki2024synthetic}, such as those arising from variations in equipment, operators, or patient populations, which enhances the objectivity and consistency. Significant advancements in synthetic medical imaging, driven by Variational Autoencoders (VAEs) \cite{doersch2016tutorial}, Generative Adversarial Networks (GANs) \cite{goodfellow2020generative}, and diffusion models \cite{ho2020denoising}, have revolutionized dataset augmentation, multi-modal imaging, and the generation and removal of pathological features.
GANs have enhanced CT and MRI data by preserving essential features while increasing variability, and diffusion models have further improved image quality ~\cite{al2023usability,khader2022medical}. Multi-modal synthesis and modality conversion enable cross-modal analysis and enhance diagnostic robustness~\cite{liu2021unified,abu2021paired}. Pathology generation and removal models, such as tumor synthesis in MRI/CT and pulmonary nodule generation in radiography, have significantly improved related model performance ~\cite{cohen2021gifsplanation, dai2024sasamim, chen2024towards}.

A GAN-based framework, BLS-GAN~\cite{wang2024bls}, was proposed integrated with imaging principles to achieve bone region generation in conventional radiography of joints, effectively eliminating overlapped regions, which offers a significant contribution to RA research. Although it successfully extracts the upper and lower bone textures, its primary focus is on the generation of bone regions, with a comparative limitation in the generation of soft tissue and layer separation between bone and soft tissue. Given that soft tissue is a critical component of joint structures, its realistic generation is essential to enhance the realism and diversity of bone region generation. Without the generation of soft tissue, this work is limited to tasks involving partially background-free quantification of JSN progression.

At present, there is a notable gap in research concerning the synthetic data for RA, and an absence of comprehensive data synthesis methods. Furthermore, traditional data augmentation with generation models faces significant limitations, particularly in their inability to effectively adjust JSW and address the challenges posed by imbalanced data distribution. However, the layer separation method employed by BLS-GAN warrants attention. Separately extracting the tissue layers of the image, followed by adjusting bones and reconstruction, can be considered as a process for constructing synthetic data for RA.

This work proposes an innovative finger joint \textbf{L}ayer \textbf{S}eparation \textbf{N}etworks, \textbf{LSN}, to achieve a high-accuracy separation of upper bone, lower bone, and soft tissue layers, with the aim of providing foundational support for more comprehensive and diverse synthetic data. Specifically, our research contributions are as follows.
\begin{itemize}
    \item \textbf{Layer separation networks}: A novel network architecture to achieve highly realistic separation of bones and soft tissue layers simultaneously. Soft tissue discrimination network and random shifting function are introduced to achieve smooth and realistic soft tissue generation.

    \item \textbf{Adjustable JSW images synthesis}: 
    A challenging scenario in synthetic data, adjustable JSW images synthesis is introduced, and a state-of-the-art image synthesis method is proposed. The synthetic images are highly realistic, providing a valuable resource for advancing research in RA.

    \item \textbf{Improvement of downstream tasks}: The synthetic joint data demonstrates the potential to significantly improve the accuracy, stability, and robustness of downstream tasks while reducing the reliance on annotated training data .
\end{itemize}

\begin{figure*}[!t]
    \centering
    \includegraphics[width=\linewidth]{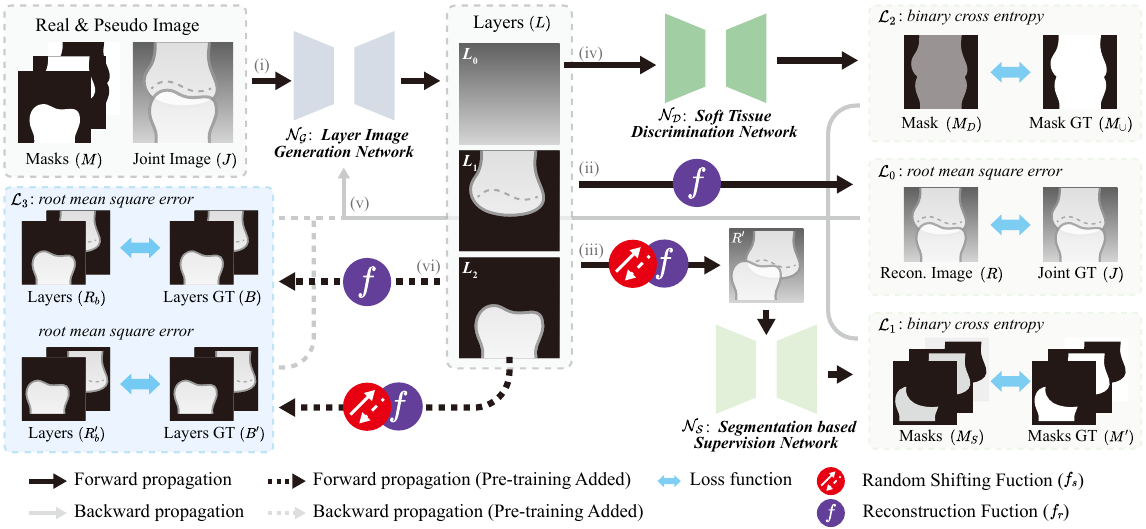}
    \caption{\textit{Layer Separation}: it generates the layer images of upper and lower bones and soft tissues based on a single image. The LSN consists of five main components: a generation network $\mathcal{N_G}$, a supervision network $\mathcal{N_S}$, a discrimination network $\mathcal{N_D}$, a random shifting function $f_s$ and a reconstruction function $f_r$. The generation process is performed as follows: 
    (i) The $\mathcal{N_G}$ processes the original joint images and the corresponding bone masks as input to generate the layer images.
    (ii) The layer images are processed through a $f_r$ to obtain a reconstruction image. 
    (iii) The layer images are processed through functions $f_r$ and $f_s$ to generate a shifted reconstruction image, which serves as input to the $\mathcal{N_S}$, yielding segmentation masks for the upper, lower bones and the soft tissue.
    (iv) The soft tissue layer image is extracted and input into the $\mathcal{N_D}$, producing a regional segmentation mask for bone shadows.
    (v) Construct a hybrid loss function including the discrepancy $\mathcal{L}_1$ between the $\mathcal{N_S}$ mask and the GT, the discrepancy $\mathcal{L}_0$ between the reconstructed image and the original image, and the dual discrepancy $\mathcal{L}_2$ between the mask from the soft tissue $\mathcal{N_D}$ and the GT. 
    (vi) The LSN training is conducted in two stages. In the first-training stage, using pseudo-images, the discrepancy $\mathcal{L}_3$ between the pseudo and reconstructed bone layers (with and without random shifting) is incorporated into the original loss function. Subsequently, a second -training stage is performed in both real and pseudo-images using the original loss function. }
\label{fig:pipeline}
\end{figure*}

\section{Methodology}

\subsection{Problem Formulation}
We propose a challenging research scenario for RA synthetic images: \textit{Adjustable JSW Images Synthesis}. 
The problem formulation in the adjustable JSW images synthesis involves (i) layer separation for soft tissue, upper and lower bones, (ii) reconstruction from the layer images after shifting (specified parameters or random parameters).
This requires addressing several critical challenges: (i) ensuring that the generated layer images adhere to the radiographic imaging principles in the overall image, (ii) eliminating the bone overlap in finger joints caused by disease progression or improper hand positioning, (iii) generating clear and homogeneous soft tissue layers that are devoid of bone shadows and conform to radiographic characteristics.

\subsection{Layer Separation Networks}
Assuming that in conventional radiography of finger joints, the joint image is formed only through the overlap of upper, lower bones and soft tissue textures, following specific principles. The LSN is proposed to extract layer images of the upper bone, lower bone, and soft tissue in conventional finger joint radiography. As shown in Fig.~\ref{fig:pipeline}, the LSN consists of three basic sub-networks: the layer image generation network, the segmentation-based supervision network, and the soft tissue discrimination network. We define the generated layer images as \(0, ..., i, ..., n\), where $n$ is set to 2 and $0$ is set as the soft tissue layer, $1$ and $2$ are set as the lower bone and upper bone layers.

\paragraph{Layer Image Generation Network}
We perform a generation network to separate the texture in the layer domain and generate images that conform to the texture distribution of each layer. The backbone network here can be any generation network; we performed transUnet \cite{chen2021transunet} here. The input is the joint image \(J\) and its corresponding masks \(M = \{M_0, ..., M_n\}\). The output of the generation network is defined as \(L = \{L_0, ..., L_n\}\). Assuming the layer image generator is denoted as \(\mathcal{N_G}\), the generation process can be defined as Eq.\ref{eq:n_generation}
\begin{equation}
L = \mathcal{N_G}(J)\cdot M
\label{eq:n_generation}
\end{equation}

\paragraph{Segmentation-based Supervision Network}
We integrate a segmentation network for pixel-level supervision. The network outputs masks with three channels containing upper and lower bones without overlapped regions, and soft tissue. Through the principle of adversarial generation, the network achieves pixel-level differentiation of soft tissue, bones, and overlapped bone region distributions in the shifted reconstruction joint image, which enables unsupervised optimization without layer-independent GT images while partially mitigating the bone shadows. The segmentation network supports the use of various backbones. In this study, Unet is employed as the backbone.

The layer images from the generation network $L$ serve as the input. The output is defined as \(M = \{M_0, ..., M_n\}\). Suppose that the segmentation network is denoted as \(\mathcal{N_S}\). Therefore, the supervision process can be defined as Eq.\ref{eq:n_supervision}, where $M_S$ represents the segmentation mask. 
\begin{equation}
M_S = \mathcal{N_S}(R')
\label{eq:n_supervision}
\end{equation}

\paragraph{Soft Tissue Discrimination Network}
In our perspective, the presence of bone shadows in the soft tissue severely affects the quality of the generated results. Thus, we introduce a discrimination network to achieve bone shadow segmentation and supervise the generation network. Our objective can be described as ensuring the bone shadow regions in the generated soft tissue images are indistinguishable, while maintaining a consistent texture distribution between bone shadow and non-bone shadow regions.
Therefore, the discrimination network is structured to produce a lower loss when bone shadows are present and a higher loss when bone shadows are absent or substantially reduced. Meanwhile, for adversarial training, the loss function of the generation network is formulated as the dual counterpart of the loss function in the discrimination network.

The soft tissue layer images $L_0$ serve as the input to the network. The output is defined as \(M_D\). Suppose that the discrimination network is denoted as \(\mathcal{N_D}\). Therefore, the discrimination process can be defined as Eq.\ref{eq:n_discrimination}.
\begin{equation}
M_D = \mathcal{N_D}(L_0)
\label{eq:n_discrimination}
\end{equation}

\paragraph{Radiography Imaging Principles based Reconstruction}
According to the principles of conventional radiography, different tissues exhibit varying absorption rates. Tissues with higher density demonstrate greater absorption, while those with lower density exhibit weaker absorption, resulting in radiographic representations ~\cite{bushberg2011essential, huda2015radiographic}. In the presence of tissue overlap, the X-ray absorption by the upper tissues influences the imaging of the lower tissues, showing an exponential decay.

Therefore, we introduce a reconstruction process for layer images. In this process, the image is reconstructed according to the reconstruction function \(f_r\), as defined in Eq.~\ref{eq:f_reconstruction}, where \( R \) denotes the reconstructed image. Specifically, the images of the absorption rate can be defined as \(1 - L\). 
\begin{equation}
R = f_r(L)= 1- \prod_{i=0}^n(1-L_i)
\label{eq:f_reconstruction}
\end{equation}

\paragraph{Random Shifting}
To remove bone shadows more accurately and introduce the synthesis process into the network architecture, we incorporate a random shifting process of bones, which serves as a supervisory mechanism for the generation network. Our generation process adheres to the principles of radiographic imaging. Ideally, given accurately generated soft tissue, reconstruction after random shifting will not introduce bone shadows, and the reconstructed images can be correctly segmented by the supervision network. 

The input element \(A\) is randomly shifted according to the function \(f_s\), as defined in Eq.~\ref{eq:transformed}, where \(A'\) represents the shifted elements and \(t\) denotes the shifting matrix with translation \(x_i, y_i\) and rotation \(\theta_i\). Therefore, the reconstructed shifted image \(R'\) and the shifted mask \(M'\) can be defined as \(R' = f_r(f_s(L, t))\), \(M' = f_s(M, t)\)
\begin{equation}
t_i = 
\begin{bmatrix}
\cos(\theta_i) & -\sin(\theta_i) & x_i \\
\sin(\theta_i) & \cos(\theta_i) & y_i \\
\end{bmatrix}
\label{eq:ti}
\end{equation}
\begin{equation}
A' = f_s(A, t) = \{ A_0, A_i \cdot t_i\ | i=1, \dots, n\}
\label{eq:transformed}
\end{equation}

\paragraph{Loss Function}
We construct the loss function based on binary cross entropy (BCE) loss \(\mathcal{L}_{b}(y, \hat{y})\)~\cite{yeung2022unified} and root mean squared error (RMSE) loss \(\mathcal{L}_{r}(y, \hat{y})\)~\cite{chai2014root}, where \(y\) represents the predicted value and \(\hat{y}\) represents the GT.

For the supervision of the generation network, the network loss function consists of three parts: reconstruction loss, supervision network loss, and soft tissue discrimination loss, which can be defined as Eq. \ref{eq:mainloss}, where \(M_\cap = \bigcap_{i=1}^{n} M_i \), \(M_\cup = \bigcup_{i=1}^{n} M_i \). According to experimental experience, the weights of each loss function are as follows: \(\alpha = 0.6, \beta = 0.3, \gamma = 0.1\).
\begin{equation}
\begin{aligned}
& \mathcal{L}_{0} = \mathcal{L}_{r}(R, J) + \mathcal{L}_{r}(R_\cap, J_\cap) \\
& \mathcal{L}_{1} = \mathcal{L}_{b}(\mathcal{N_S}(R'), M') \\
& \mathcal{L}_{2} = 1 - \mathcal{L}_{b}(\mathcal{N_D}(L_0), M_\cup) \\
& \mathcal{L} = \alpha \mathcal{L}_{0} + 
\beta \mathcal{L}_{1} + \gamma \mathcal{L}_{2}
\end{aligned}
\label{eq:mainloss}
\end{equation}

In addition, we train the supervision and discrimination networks simultaneously and independently. 
The input of the supervision network is the original image \(J\), and the corresponding GT is the masks without overlapped regions, denoted as \(M' =\{M_1 - M_\cap, ..., M_n - M_\cap \}\).
The loss function is defined as Eq.~\ref{eq:segloss}.
\begin{equation}
\mathcal{L}_S = \mathcal{L}_b(\mathcal{N_D}(J), M')
\label{eq:segloss}
\end{equation}

As for the soft tissue discrimination network, the input of the network is the generated soft tissue layer image \(L_0\), and the corresponding GT is the union of bone masks \(M_\cup \). The loss function is defined as Eq.~\ref{eq:disloss}.
\begin{equation}
\mathcal{L}_D = \mathcal{L}_b(\mathcal{N_D}(L_0), M_\cup)
\label{eq:disloss}
\end{equation}

\paragraph{Pseudo Images and Two-stage training}
In order to improve the stability and accuracy of the network, we create pseudo-images $\tilde{J}$ for two-stage training, with overlapped regions based on non-overlapped images by modifying the image processing described in BLS-GAN~\cite{wang2024bls}. In particular, the correction parameter $k$ is determined by solving the Laplace equation, after which the shifted upper and lower bones are reconstructed, and the soft tissue region is subsequently spliced, defined as Eq.~\ref{eq:pseudoimage}, where $J'$ and $M'$ represent the image and the corresponding masks with random scaling and translation. \(B = \{B_i = J'\cdot M'_i | i=1, ..., n\}\)represents the bone region with soft tissue texture.
\begin{equation}
\tilde{J} = (1- k\prod_{i=1}^n(1- B_i)) + J' \cdot \bigcup_{i=1}^{n}M'_i
\label{eq:pseudoimage}
\end{equation}

We perform the first stage using pseudo-images as dataset \(\mathcal{D}_1\). Specifically, since the pseudo-images are created from non-overlap images, we can effectively obtain non-overlap upper and lower bone GT. Therefore, $B$ and the randomly shifted $B' = f_s(B, t_b)$ are included as GT, and the modified loss function is denoted as Eq. \ref{eq:pregen}, 
where \(R_b = f_r(L) \cdot M_b\),  \(M_b = \{M_1, ..., M_n\}\), \(R'_b = f_r(f_s(L, t_b) \cdot M'_b\), \(M'_b = f_s(M_b, t_b)\).
In addition, when training the segmentation network, $J$ (non-overlap) is also used as a non-overlap image sample for the loss function, denoted as Eq.\ref{eq:preseg}. The weights of the loss functions are as follows: \(\alpha' = 0.5, \beta' = 0.2, \gamma' = 0.2, \delta = 0.1; \alpha'' = 1, \beta'' = 0.4, \delta' = 0.4\).
\begin{equation}
\begin{aligned}
& \mathcal{L}_{3} = 0.5\times \mathcal{L}_r(R_b, B) + 0.5\times \mathcal{L}_r(R'_b, B') \\
& \tilde{\mathcal{L}}  =
\alpha' \mathcal{L}_{0} + 
\beta' \mathcal{L}_{1} + \gamma' \mathcal{L}_{2} + \delta \mathcal{L}_{3}, \text{if } \text{epoch} > m \\
& \tilde{\tilde{\mathcal{L}}}  =
\alpha'' \mathcal{L}_{0} + \beta'' \mathcal{L}_{1} + \delta \mathcal{L}_{3}, \text{if } \text{epoch} \leq m
\end{aligned}
\label{eq:pregen}
\end{equation}
\begin{equation}
\tilde{\mathcal{L}_S} = 0.5\times\mathcal{L}_b(\mathcal{D}(\tilde{J}), M') +  0.5\times\mathcal{L}_b(\mathcal{D}(J), M)
\label{eq:preseg}
\end{equation}
In the second training, due to the absence of GT for the upper and lower bones in real images, we continue to apply the original loss function and utilize both pseudo and real images as a dataset \(\mathcal{D}_2\). Therefore, the training pipeline can be defined as Algorithm \ref{al:train}.

\begin{algorithm}[tb]
\caption{LSN Training Process}
\begin{algorithmic}
\STATE \textbf{Input:} Training dataset $\mathcal{D}_1$, $\mathcal{D}_2$, learning rates $\eta_\mathcal{G}, \eta_\mathcal{S}, \eta_\mathcal{D}$, initial parameters $\theta_\mathcal{G}, \theta_\mathcal{S}, \theta_\mathcal{D}$.
\STATE \textbf{Output:} Optimized parameters $\theta_\mathcal{G}^*, \theta_\mathcal{S}^*, \theta_\mathcal{D}^*$.

\STATE \textbf{\textit{Stage 1: Train LSN in $\mathcal{D}_1$}}
\FOR{$epoch$}
    \STATE $\theta_\mathcal{G}^* \gets \theta_\mathcal{G} - \eta_\mathcal{G} \nabla_{\theta_\mathcal{G}} \tilde{\tilde{\mathcal{L}}}$
    \STATE $\theta_\mathcal{S}^* \gets \theta_\mathcal{S} - \eta_\mathcal{S} \nabla_{\theta_\mathcal{S}} \tilde{\mathcal{L}_S}$
    \IF{$epoch > m$}
        \STATE $\theta_\mathcal{G}^* \gets \theta_\mathcal{G} - \eta_\mathcal{G} \nabla_{\theta_\mathcal{G}} \tilde{\mathcal{L}}$
        \STATE $\theta_\mathcal{D}^* \gets \theta_\mathcal{D} - \eta_\mathcal{D} \nabla_{\theta_\mathcal{D}} \mathcal{L}_{D}$
    \ENDIF
\ENDFOR

\STATE \textbf{\textit{Stage 2: Train LSN in $\mathcal{D}_2$}}
\FOR{$epoch$}
    \STATE $\theta_\mathcal{G}^* \gets \theta_\mathcal{G} - \eta_\mathcal{G} \nabla_{\theta_\mathcal{G}} \mathcal{L}$
    \STATE $\theta_\mathcal{S}^* \gets \theta_\mathcal{S} - \eta_\mathcal{S} \nabla_{\theta_\mathcal{S}} \mathcal{L}_S$
    \STATE $\theta_\mathcal{D}^* \gets \theta_\mathcal{D} - \eta_\mathcal{D} \nabla_{\theta_\mathcal{D}} \mathcal{L}_D$
\ENDFOR
\end{algorithmic}
\label{al:train}
\end{algorithm}

\paragraph{Implementation}
The networks were implemented on a workstation with three GPUs (NVIDIA GeForce GTX 2080 Ti). The generation, supervision, and discrimination networks were trained using the AdamW optimizer with an initial learning rate as follows: \(\eta_\mathcal{G}=1e^{-4}, \eta_\mathcal{S}=1e^{-4}, \eta_\mathcal{D}=5e^{-4}\), decreasing by 0.5 every 100 epochs.
The image size processed by LSN is set to 256 $\times$ 256. 
In our practice, we commence by performing first-stage training on \(\mathcal{D}_1\), extending this preparatory phase across 300 epochs and $m$ is set to 200, with a batch size of 12. Subsequently, we refine the loss function and GT, maintaining the same batch size, for an additional 100 epochs on \(\mathcal{D}_2\) to optimize the performance of our networks. 

\subsection{Adjustable JSW Image Synthesis}
Using the layer images generated from LSN, the adjustable JSW synthetic image $J^*$ can be efficiently achieved. The process is as follows: (i) By applying the shifting function $f_s$ with specified or random parameters in a predefined range to the upper and lower bones; (ii) Reconstructing the shifted bone layers with the soft tissue layer by $f_r$. A substantial dataset of synthetic images with varying JSW can be created from a single input image, as illustrated in Eq.\ref{eq:sysnthesis}, where $t^*$ denotes the shifting parameters. Combined with the original annotations (e.g., JSW and SvdH), the shifting parameters can be used to produce GT of this synthetic image.
\begin{equation}
J^* = f_r(f_s(L, t^*))
\label{eq:sysnthesis}
\end{equation}

\section{Experiments}
\subsection{Joint Image Dataset} 
The original real joint image dataset in BLS-GAN~\cite{wang2024bls} was used for training and testing. The dataset contains 430 MCP joints for 1,594 joint images with corresponding annotated bone masks, which are divided into training and testing sets at a 3:1 ratio by joint.

For downstream tasks, the JSW of each image was annotated using the method developed based on the layer separation, under the guidance of experienced radiologists. Specifically, the method enabled manual alignment of the upper and lower bones by adjusting their positions to align the boundaries of the joint contact surfaces, where the JSW value was defined as zero. Thus, the JSW was subsequently calculated as the difference between the displacements of the bone layers. The annotated JSW data will be made publicly available.
Meanwhile, we constructed the SvdH-like score annotations based on JSW annotations according to the SvdH scoring definition~\cite{van2000read,van1999reliability}. Specifically, the annotations were established according to the following criteria: 0: normal; 1: 100$\%$–75$\%$ of normal JSW; 2: 75$\%$–50$\%$ of normal JSW; 3: 50$\%$–25$\%$ of normal JSW; 4: less than 25$\%$ of normal JSW (normal average joint space: 1.75 mm ~\cite{pfeil2007computer}).

\subsection{Reconstruction Images Evaluation}
Due to the absence of GT for layer images, the evaluation was conducted exclusively on reconstructed images and real images. The network predicted the upper bone, lower bone, and soft tissue layers, subsequently reconstructing the images through the reconstruction function, which were compared to the corresponding real images for evaluation. The performance assessment employed four quantitative metrics: mean squared error (MSE), structural similarity index (SSIM), peak signal-to-noise ratio (PSNR), and Fréchet Inception Distance (FID).

\begin{figure}[!t]
  \includegraphics[width=\linewidth]{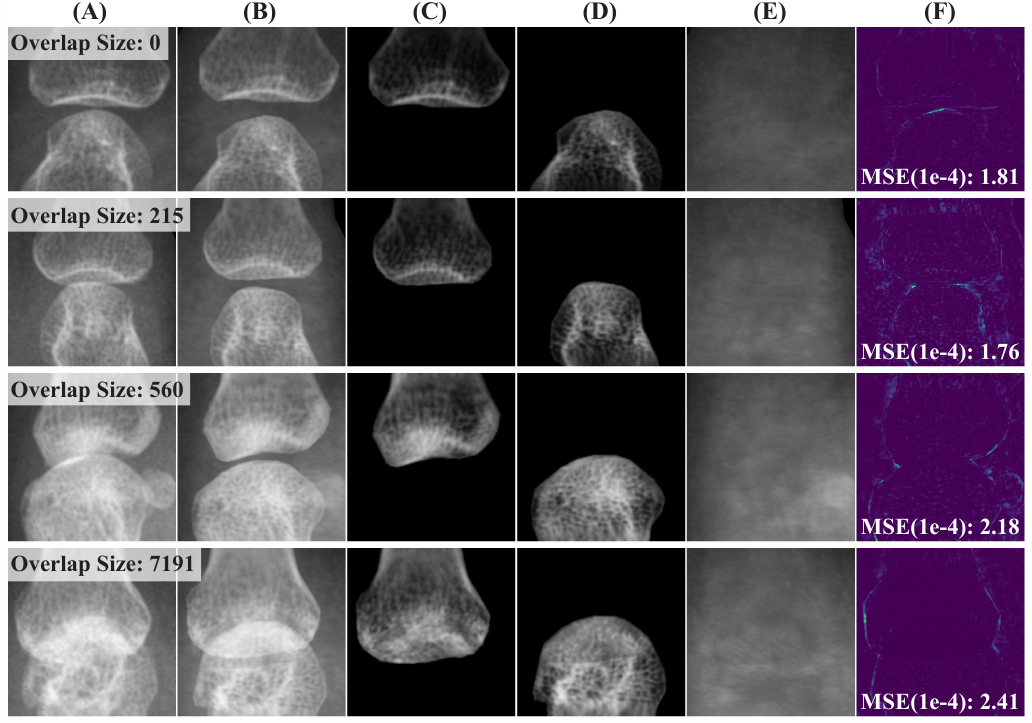}
  \centering
  \caption{Visualization results of our ablation study. (A) Real Joint image; (B) Shifted Reconstruction Joint Image; (C) Upper Bone Layer; (D) Lower bone layer; (E) Soft Tissue Layer; (F) MSE Spectrum (Reconstruction Joint Image v.s. A).}
\label{fig:evaluation}
\end{figure}

\begin{table}[!t]
\caption{Evaluation result of the proposed LSN in different metrics. Expressed as mean $\pm$ standard deviation.}
\centering
\resizebox{\linewidth}{!}{
    \begin{tabular}{lcccc}
        \toprule
        Joint & MSE (\(10^{-4}\)) & SSIM (\(10^{-2}\)) & PSNR & FID (\(10^{-2}\)) \\
        \midrule
        Thumb & 2.38 $\pm$ 0.50 & 94.91 $\pm$ 0.36 & 36.32 $\pm$ 0.86 & 3.24 $\pm$ 0.73 \\
        Index & 2.19 $\pm$ 0.31 & 94.98 $\pm$ 0.33 & 36.64 $\pm$ 0.62 & 3.03 $\pm$ 0.50 \\
        Middle & 2.20 $\pm$ 0.30 & 95.02 $\pm$ 0.38 & 36.62 $\pm$ 0.59 & 3.06 $\pm$ 0.45 \\
        Ring & 2.14 $\pm$ 0.35 & 95.07 $\pm$ 0.43 & 36.76 $\pm$ 0.68 & 2.97 $\pm$ 0.56 \\
        Small & 2.12 $\pm$ 0.32 & 95.07 $\pm$ 0.34 & 36.79 $\pm$ 0.64 & 2.95 $\pm$ 0.53 \\
        \midrule
        Overall & 2.19 $\pm$ 0.34 & 95.02 $\pm$ 0.37 & 36.66 $\pm$ 0.66 & 3.03 $\pm$ 0.53 \\
        \bottomrule
    \end{tabular}
}
\label{tab:evalutaion}
\vskip -0.1in
\end{table}

As illustrated in Table \ref{tab:evalutaion}, LSN demonstrated strong performance across various finger joints, showing exceptional stability and adaptability, and consistently produced high-quality generation outcomes. Figure \ref{fig:evaluation} further underscored the ability of LSN to generate accurate and clear layer images for both bones and soft tissue. Even under challenging conditions involving bone overlap, especially with large overlap sizes, the method effectively reconstructed the upper and lower bone layers while eliminating overlaps. Furthermore, the soft tissue layer generated by the network was characterized by a uniform texture devoid of bone shadows. In particular, the shifted reconstruction images closely resembled the real images and were free of bone shadows. Overall, the proposed method achieved accurate layer separation from single-joint images, providing a solid data foundation for generating synthetic images.

\subsection{Ablation Study}
An ablation study was performed to evaluate the impact of individual sub-networks on the overall network performance. This study involved a stepwise evaluation of various pipeline configurations, including the supervision network $\mathcal{N_S}$, discrimination network $\mathcal{N_D}$, and the first-stage training $T_1$, using the generation network $\mathcal{N_G}$ and reconstruction function $f_r$ as the $Baseline$. We also conducted a specific study for the random shifting function $f_s$, given its extensive application in multiple processes in the architecture.

\begin{figure}[!t]
  \includegraphics[width=\linewidth]{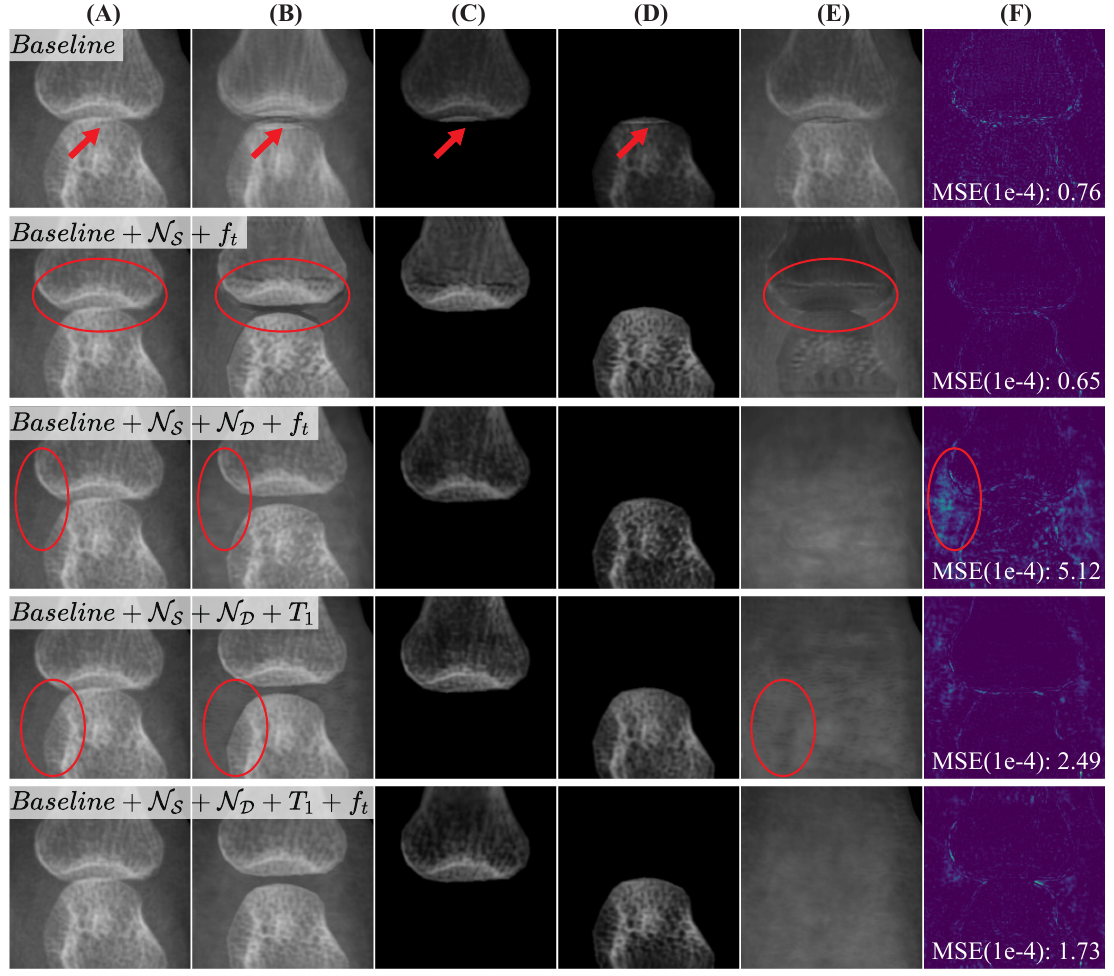}
  \centering
  \caption{Visualization results of our ablation study. (A) Real Joint image; (B) Shifted Reconstruction Joint Image; (C) Upper Bone Layer; (D) Lower Bone Layer; (E) Soft Tissue Layer; (F) MSE Spectrum (Reconstruction Joint Image v.s. A).}
\label{fig:ablation_image}
\end{figure}

\begin{table}[!t]
\caption{Comparison results in ablation study. Expressed as mean $\pm$ standard deviation.}
\centering
\resizebox{\linewidth}{!}{
    \begin{tabular}{cccccccc}
        \toprule
        $\mathcal{N_S}$ & $\mathcal{N_D}$ & $T_1$ & $f_s$ & MSE (\(10^{-4}\)) & SSIM (\(10^{-2}\)) & PSNR & FID (\(10^{-2}\)) \\
        \midrule
        & & & & 0.76 $\pm$ 0.11 & 97.98 $\pm$ 0.22 & 41.22 $\pm$ 0.53 & 1.30 $\pm$ 0.20 \\
        
        $\surd$ & & & $\surd$ & 0.77 $\pm$ 0.11 & 97.48 $\pm$ 0.17 & 41.16 $\pm$ 0.59 & 1.12 $\pm$ 0.20 \\

        $\surd$ & $\surd$ & & $\surd$ & 5.33 $\pm$ 0.88 & 93.08 $\pm$ 0.84 & 32.79 $\pm$ 0.72 & 8.77 $\pm$ 2.08 \\
        
        $\surd$ & $\surd$ & $\surd$ & & 2.80 $\pm$ 0.44 & 95.11 $\pm$ 0.48 & 35.58 $\pm$ 0.66 & 4.32 $\pm$ 0.86 \\
        
        $\surd$ & $\surd$ & $\surd$ & $\surd$ & 2.19 $\pm$ 0.34 & 95.02 $\pm$ 0.37 & 36.66 $\pm$ 0.66 & 3.03 $\pm$ 0.53 \\

        \bottomrule
    \end{tabular}
    }
\begin{tablenotes}
\item $T_1$: the first stage in pseudo-images two-stage training
\end{tablenotes}
\label{tab:ablation}
\vskip -0.1in
\end{table}

The results presented in Table \ref{tab:ablation} and Figure \ref{fig:ablation_image} demonstrated that the sub-networks, functions, and training strategies integrated into the LSN collectively contributed to and enhanced the final results. Specifically, the incorporation of $\mathcal{N_S}$ effectively guided the generation network in the absence of bone layer GT and under conditions of random shifting, thereby partially mitigating the appearance of bone shadows in the soft tissue layer. The inclusion of $\mathcal{N_D}$ substantially reduced bone shadows in the soft tissue layer, although it introduced a trade-off by slightly reducing the accuracy of the reconstructed images. The incorporation of $T_1$ significantly enhanced both accuracy and stability. The introduction of $f_s$ enabled effective supervision of texture generation in each layer, which was essential for suppressing bone shadows at the edges of the original bone region in the soft tissue layer. This also facilitated the generation of soft tissue layers with uniform texture distribution and enhanced realism.
Additionally, while excluding $\mathcal{N_D}$ resulted in higher quantitative metrics, the generated soft tissue layers contained pronounced bone shadows, rendering the results clinically unacceptable. In contrast, the inclusion of $\mathcal{N_D}$ produced more homogeneous and clinically viable soft tissue layers, although with slightly lower metric scores, which aligned with the main objectives of this study.

\subsection{Visual Turing Test}
\begin{figure}[!t]
  \includegraphics[width=0.9\linewidth]{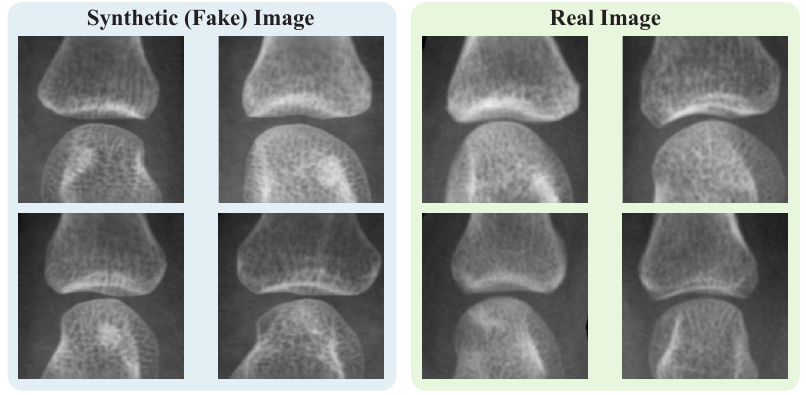}
  \centering
  \caption{Original and Synthetic images in Visual Turing Test.}
\label{fig:turing}
\end{figure}

We conducted a visual Turing test on a set of 100 images (real and synthetic images in a 1:1 ratio), as shown in Fig.~\ref{fig:turing}, which was explained to the subjects. Five subjects with 13, 18, 20, 27, and 32 years of experience as radiological technologists participated in the test.

As shown in Table~\ref{tab: turingtest}, the results of the visual Turing test indicated that the experts demonstrated moderate proficiency in distinguishing between real and synthetic images, achieving an average accuracy of approximately 0.7. However, substantial variability was observed among individual experts. In particular, R1 and R3 exhibited high sensitivity and specificity, while R5 performed poorly in all metrics, underscoring the challenges associated with recognizing synthetic images. These results suggest that the generated images effectively replicate the characteristics of real images to a certain extent, exhibiting a similar texture distribution. Consequently, even experienced experts face considerable difficulty in differentiating between synthetic and real images.
Compared to Wang et al.~(\citeyear{wang2024bls}), our task is more complex and difficult, which is primarily due to the additional generation of soft tissue and synthesis steps, including the random shifting of bones and reconstruction. 

\begin{table}[!t]
\caption{Visual Turing Test evaluation results across three image groups. Radiological technologists were tasked with labeling each set of images as real or fake.}
\centering
\resizebox{0.9\linewidth}{!}{
    \begin{tabular}{lcccccc}
        \toprule
        & R1 & R2 & R3 & R4 & R5 & Overall \\
        \midrule
        sensitivity & 0.82 & 0.72 & 0.78 & 0.64 & 0.56 & 0.70  \\
        specificity & 0.60 & 0.86 & 0.78 & 0.76 & 0.56 & 0.71 \\
        accuracy & 0.71 & 0.79 & 0.78 & 0.70 & 0.56 & 0.71 \\
        \bottomrule
    \end{tabular}
}
\label{tab: turingtest}
\vskip -0.1in
\end{table}

\begin{table*}[t!]
\caption{Downstream Tasks Evaluation result of w/ (pre-training pipeline) and w/o (control pipeline) synthetic data pre-training from the LSN in different metrics. In the synthetic data, based on the original image, 8 times synthesis is performed.}
\centering
\resizebox{\linewidth}{!}{
    \begin{tabular}{c|ccc|cccc|cccc}
    \toprule
    \multirow{2}{*}{Synthetic Data} & \multicolumn{3}{c|}{JSN Progress} & \multicolumn{4}{c|}{JSW} & \multicolumn{4}{c}{SvDH-like JSN Score} \\ 
    \cline{2-12}
     & MSE ($10^{-3}$) & $\sigma$ ($10^{-4}$) & $\sigma'$ ($10^{-3}$) & MSE ($pi^2$) & MAE ($pi$) & EVS & $R^2$ & ACC & SEN & SPC & PRE \\
    \midrule
    w/o & 2.7225 $\pm$ 1.2628 & 12.4007 & 1.3999 
     & 8.3166 & 1.8138 & 0.5862 & 0.5830 
     & 0.8628 & 0.4658 & 0.8910 & 0.5321 \\ 
    w/ & \textbf{2.2545} $\pm$ \textbf{1.1155} & \textbf{8.3340} & \textbf{1.1910}
    & \textbf{4.6437} & \textbf{1.5500} & \textbf{0.7689} & \textbf{0.7672} 
    & \textbf{0.8954} & \textbf{0.5870} & \textbf{0.9167} & \textbf{0.6949} \\
    \bottomrule
    \end{tabular}
}
\label{tab:downstream}
\vskip -0.1in
\end{table*}

\subsection{Improvement in Downstream Tasks}
We verified the improvement of the model in downstream tasks by introducing LSN synthetic data pre-training in controlled experiments (\textit{Pre-training pipeline: downstream model + pre-training; Control pipeline: downstream model}), where synthetic data is created in multiples from the original images, and the GT is automatically generated by the LSN. Downstream tasks include JSN progress quantification, JSW quantification, and SvdH-like JSN score qualification. The networks in the experiment were trained to full convergence with a uniform total number of epochs.

\paragraph{JSN Progress Quantification}
For the JSN progress quantification, we performed the deep registration method proposed in ~\cite{wang2023deep}. The evaluation of the experimental results followed the metrics established in \cite{ou2022sub}, including the mean squared error (MSE), standard deviation ($\sigma$), and phase standard deviation ($\sigma'$).

As shown in Table \ref{tab:downstream}, the pre-training pipeline outperformed the control pipeline, achieving lower values across all three evaluation metrics. The reduction in MSE reflected a modest improvement in accuracy. Moreover, the decreases in the two standard deviation metrics ($\sigma$, $\sigma'$) indicated substantial enhancements in stability and robustness. These experimental results aligned with the expected performance gains associated with the pre-training, demonstrating its effectiveness in improving both accuracy and consistency.

\paragraph{JSW Quantification}
For the JSW quantification, the conventional method typically employs a supervised edge detection algorithm combined with edge distance calculation, which heavily depends on manual annotation. Therefore, the experiments conducted a ResNet50-based regression network for JSW quantification. The evaluation of outcomes was conducted using metrics including mean squared error (MSE), mean absolute error (MAE), explained variance score (EVS), and the coefficient of determination ($R^2$).

As presented in Table \ref{tab:downstream}, the overall accuracy of JSW quantification using the regression method remained relatively low, due to the abandonment of edge detection, which was consistent with historical observations in practical applications. However, the pre-trained pipeline demonstrated significantly enhanced performance across multiple evaluation metrics compared to the control pipeline. These improvements were evident in all aspects of the evaluation, underscoring the effectiveness of LSN synthetic image pre-training in substantially improving the accuracy, stability, and robustness of the JSW measurement model.

\paragraph{SvdH-like JSN Score Qualification}
For the SvdH-like JSN score qualification, a ResNet-50-based classification network was employed as the downstream task network for evaluation. The experimental results were evaluated using several performance metrics, including accuracy (ACC), sensitivity (SEN), specificity (SPC), and precision (PRE).

The SvdH scoring system, as a qualification system based on human visual assessment, often involves redundancy and ambiguity. Consequently, a straightforward division and construction of an SvdH-like scoring system based on JSW could not fully and accurately simulate the original SvdH system. This limitation was reflected in the overall accuracy, which remains below 0.95. Nonetheless, in the SvdH-like scoring system, as presented in Table \ref{tab:downstream}, the integration of LSN synthetic images pre-training has led to notable improvements across multiple evaluation metrics compared to the control pipeline. These enhancements demonstrate the effectiveness of the pre-training pipeline in boosting the accuracy, stability, and robustness of the model.

\begin{figure}[!t]
  \includegraphics[width=\linewidth]{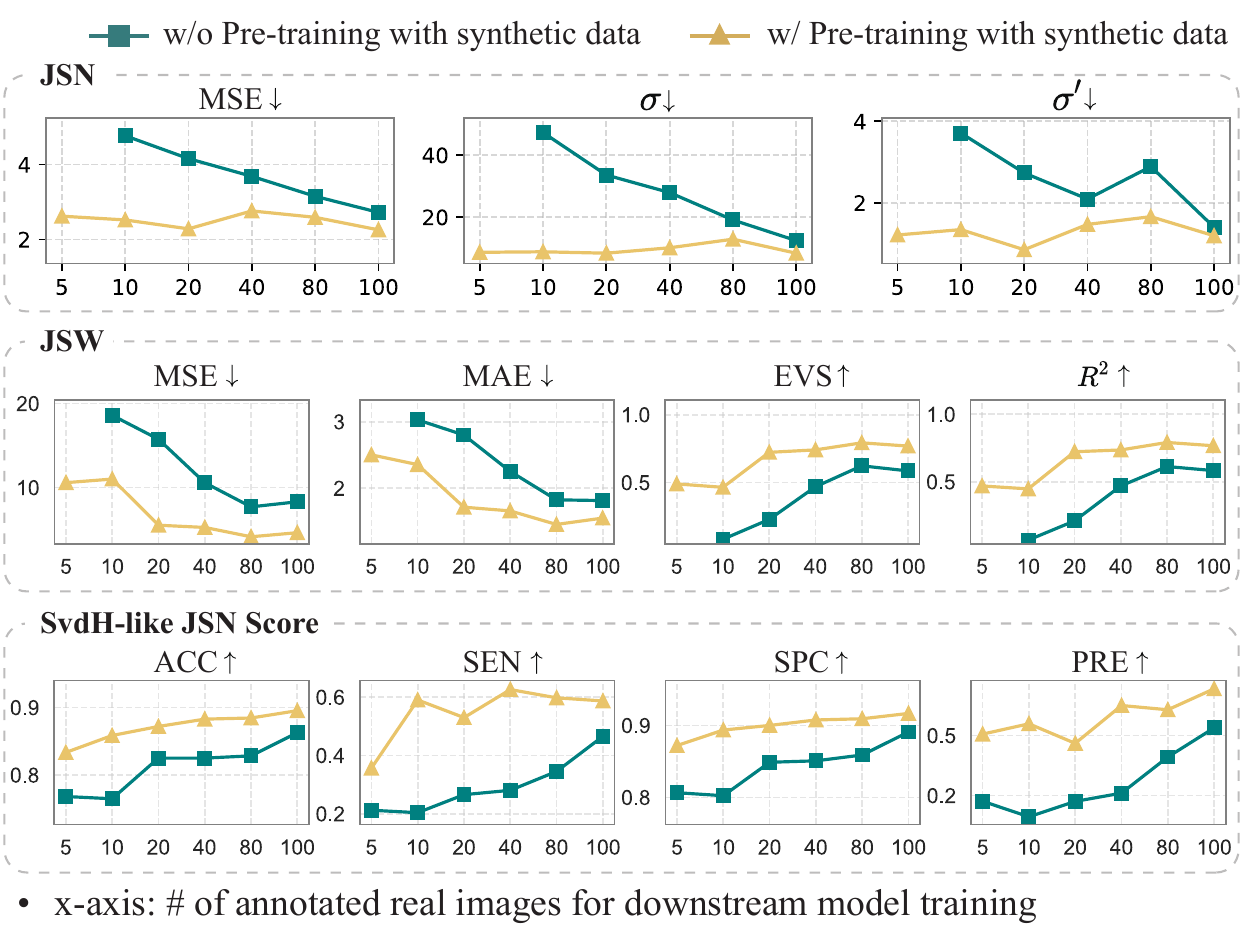}
  \vskip -0.1in
  \centering
  \caption{Reduced annotations for the downstream models.}
\label{fig: reduce}
\end{figure}

\paragraph{Reduced Annotations for Training}
The performance of downstream models in registration, regression, and classification is highly dependent on the amount and diversity of annotated training data used during the training stage ~\cite{jaipuria2020deflating}. Therefore, we studied the relationship between the amount of real annotated images and the performance of downstream task models in controlled experiments. Specifically, for real annotated data, we reduced it by a certain ratio (5$\%$, 10$\%$, 20$\%$, 40$\%$, 80$\%$ of the original datasets), and for synthetic data, we reduced the original real data used to create synthetic data and amplified it into an equal amount of synthetic data.

As illustrated in Figure \ref{fig: reduce}, the control pipeline exhibits a linear correlation between performance and the number of annotated training data, highlighting the critical role of the influence of the number of annotated training data on the final results. With limited training data, the network demonstrated issues such as instability, overfitting, and susceptibility to local optimization. Notably, tasks trained with a minimal amount of annotated data (e.g., 5$\%$) showed pronounced overfitting, and the model failed to converge in some cases. In contrast, the pre-training pipeline achieved significantly greater stability and accuracy under the same conditions. Even with a limited amount of annotated data (e.g., 5$\%$ and 10$\%$), the network demonstrated relatively stable performance and maintained its ability to train and fine-tune effectively. The results underscore that incorporating synthetic data pre-training significantly enhanced the robustness and reliability of the downstream models, leading to more stable and accurate outcomes. It also reduced the dependence on high-precision, manually annotated data.

\section{Conclusion}
In this study, the Layer Separation Networks is proposed and implemented for layer separation and adjustable JSW image synthesis, which effectively addresses existing challenges in data distribution and variety, and achieves the generation of GT annotation.
Experimental results demonstrated that LSN generates high-quality images of the upper,  lower bone, and soft tissue layer images, aligning with imaging principles. Furthermore, the reconstructed images exhibit a high similarity to the original images and perform well in the Turing test. Additionally, evaluation of downstream tasks indicated that synthetic data pre-training significantly enhanced the robustness, stability and accuracy of downstream models. This study can provide a valuable support for RA-related research and CAD development.

\section*{Software and Data}
If a paper is accepted, the code and dataset will be publicly available with the camera-ready version of the paper whenever appropriate.

\section*{Impact Statement}
This paper presents work whose goal is to advance the field of 
Machine Learning. There are many potential societal consequences 
of our work, none which we feel must be specifically highlighted here.

\bibliographystyle{icml2025}
\bibliography{reference}

\end{document}